\documentclass{styles/svproc}
\usepackage{url}
\usepackage{amsmath}
\usepackage{graphicx}
\usepackage{array}
\usepackage{booktabs}
\usepackage{subcaption}

\usepackage[backend=biber, sorting=none]{biblatex}
\usepackage{multirow}

\begin{document}
\mainmatter              
\title{COBRA: Multimodal Sensing Deep Learning Framework for Remote Chronic Obesity Management via Wrist-Worn Activity Monitoring}
\titlerunning{COBRA: Chronic Obesity Behavioral Recognition Architecture}  
%
\author{Zhengyang Shen\inst{1}, Bo Gao\inst{1}, Mayue Shi\inst{1,2*} }
\authorrunning{Zhengyang Shen, Bo Gao, Mayue Shi} 
%
%
\institute{Department of Electrical and Electronic Engineering, Imperial College London, London SW7 2AZ, UK,\\
\and
Institute of Biomedical Engineering, Department of Engineering Science, University of Oxford, Oxford OX3 7DQ, UK\\
\text{* E-mail: } \email{m.shi16@imperial.ac.uk}
}

\maketitle              

\begin{abstract}

Chronic obesity management requires continuous monitoring of energy balance behaviors, yet traditional self-reported methods suffer from significant underreporting and recall bias, and difficulty in integration with modern digital health systems. This study presents COBRA (Chronic Obesity Behavioral Recognition Architecture), a novel deep learning framework for objective behavioral monitoring using wrist-worn multimodal sensors. COBRA integrates a hybrid D-Net architecture combining U-Net spatial modeling, multi-head self-attention mechanisms, and BiLSTM temporal processing to classify daily activities into four obesity-relevant categories: Food Intake, Physical Activity, Sedentary Behavior, and Daily Living. Validated on the WISDM-Smart dataset with 51 subjects performing 18 activities, COBRA's optimal preprocessing strategy combines spectral-temporal feature extraction, achieving high performance across multiple architectures. D-Net demonstrates 96.86\% overall accuracy with category-specific F1-scores of 98.55\% (Physical Activity), 95.53\% (Food Intake), 94.63\% (Sedentary Behavior), and 98.68\% (Daily Living), outperforming state-of-the-art baselines by 1.18\% in accuracy. The framework shows robust generalizability with low
demographic variance ($<$3\%), enabling scalable deployment for personalized obesity interventions and continuous lifestyle monitoring.

\end{abstract}

\section{Introduction}

Obesity has emerged as one of the most pressing global health challenges of the 21st century, with adult obesity rates more than doubling since 1990 and affecting one in eight people worldwide by 2022 \cite{WHOobesity2022website}. This alarming trend has significant implications for healthcare systems, as obesity is a major risk factor for type 2 diabetes, cardiovascular disease, and various forms of cancer \cite{ElmalehSachs2023}. Research demonstrates that even modest weight reductions of 5–10\% can yield improvements in metabolic health and quality of life \cite{ross2009future}. However, achieving sustainable weight loss remains challenging, with traditional approaches often failing to provide continuous, personalized support necessary for long-term success. The complexity of chronic obesity management stems from its multifactorial nature, involving intricate interactions between food intake, physical activity, sedentary behavior, and lifestyle patterns that require comprehensive monitoring for effective intervention.

Traditional monitoring approaches, including food diaries and activity logs, are inherently limited by their reliance on self-reporting and subjective recall, with studies showing dietary intake is often underestimated by 20-40\% \cite{thomaz2015practical}. Early wearable-based activity recognition systems primarily employed handcrafted feature extraction followed by classical machine learning algorithms. Ramachandran et al. demonstrated that combining inertial measurement unit data with vital signs significantly improved fall detection accuracy to 93\% using Random Forest \cite{ramachandran2020evaluation}, while Mondol and Stankovic developed HAWAD for hand washing detection, effective improvement in F1-score compared to the baseline method through Mahalanobis distance filtering \cite{mondol2020hawad}. These traditional methods established foundational principles but remained limited by their reliance on manually engineered features and inability to capture complex, non-linear behavioral patterns.

The advent of deep learning has revolutionized human activity recognition by enabling automatic feature extraction and capturing complex temporal dependencies. Recent advances have focused on specialized architectures for time-series analysis, with convolutional neural networks proving effective for spatial feature extraction and recurrent neural networks excelling at temporal modeling. Mekruksavanich and Jantawong proposed ResNeXt achieving 98.89\% accuracy with only 24,563 parameters \cite{mekruksavanich2022deep}, while Gupta developed a CNN-GRU hybrid model achieving 96.54\% accuracy on smartwatch data \cite{gupta2021deep}. The integration of attention mechanisms has further advanced performance, with Khatun et al. introducing CNN-LSTM with self-attention achieving 99.93\% accuracy \cite{khatun2022deep}, and Dirgová Luptáková et al. pioneering Transformer applications to achieve 99.2\% accuracy on the KU-HAR dataset \cite{dirgova2022wearable}. More sophisticated hybrid architectures have emerged, such as Li and Wang's ResNet-BiLSTM achieving over 97\% accuracy across multiple datasets \cite{li2022human}, and Mim et al.'s GRU-INC combining GRU with attention and Inception modules achieving F1-scores exceeding 96\% \cite{mim2023gru}.

Wearable digital health technologies for obesity management have demonstrated clinical efficacy through multiple applications. Weiss et al. developed Actitracker achieving over 95\% accuracy for personalized activity recognition \cite{weiss2016actitracker}, while Zhu et al. pioneered deep learning for energy expenditure estimation, achieving 30-35\% lower error rates compared to traditional methods \cite{zhu2015using}. Clinical trials have validated wearable intervention effectiveness, with randomized controlled studies showing significant BMI reduction using wrist-worn activity trackers \cite{biscuit2017shortterm} and 4.4 kg weight loss over 6 months when combining wearable devices with behavioral support \cite{cadmus2016technology}. A umbrella review in 2024 analyzing 51 systematic reviews found that wearable devices increase physical activity by a median of 1,312 steps per day \cite{umbrella2024wearable}, while network meta-analysis demonstrated effective body weight and BMI reduction in individuals with overweight/obesity \cite{bjsm2021health}.

Eating behavior detection represents one of the most challenging aspects of wearable-based obesity monitoring due to subtle and context-dependent consumption patterns. Dong et al. pioneered wrist motion pattern tracking for eating period detection, achieving 81\% accuracy in free-living conditions \cite{dong2013detecting}, while Thomaz et al. developed practical eating moment recognition using commodity smartwatches with F-scores reaching 76.1\% \cite{thomaz2015practical}. Advanced approaches have employed sophisticated deep learning methods, with Kyritsis et al. modeling wrist micromovements through CNN-based detection and LSTM temporal modeling, achieving F1-scores of 0.913 for food intake cycle detection \cite{kyritsis2019modeling}. Multi-modal approaches have shown promise, with Mirtchouk et al. combining motion sensors and audio to achieve 92\% weighted accuracy \cite{mirtchouk2017recognizing}, and specialized devices like Farooq and Sazonov's eyeglasses-based system achieving 99.85\% F1-score for multiclass food intake classification \cite{farooq2016novel}.

Despite significant advances, current approaches face critical limitations that motivate more sophisticated solutions. Most existing systems focus on general activity recognition rather than clinically relevant behavioral categories specific to obesity management, creating substantial gaps between technical performance and clinical utility. Current systems typically provide coarse-grained classifications lacking specificity for personalized obesity interventions, with limited ability to distinguish between different eating behaviors or sedentary activities crucial for comprehensive management. Many approaches demonstrate performance on small, homogeneous datasets but fail to maintain accuracy across diverse demographic groups, raising scalability concerns for population-level deployment.

This work presents COBRA (Chronic Obesity Behavioral Recognition Architecture), a comprehensive deep learning framework specifically designed to address these limitations through novel hybrid neural network architecture combining U-Net spatial modeling, multi-head self-attention mechanisms, and bidirectional LSTM networks. COBRA employs a clinically oriented classification scheme categorizing behaviors into four key domains: Food Intake, Physical Activity, Sedentary Behavior, and Daily Living. Our experimental validation on the WISDM-Smart dataset demonstrates that COBRA achieves 96.86\% overall accuracy with F1-scores exceeding 94\% across all behavioral categories, showing robust generalizability with less than 3\% performance variance across demographic subgroups. The framework represents a significant advancement toward practical, scalable solutions for continuous behavioral monitoring in chronic obesity management, providing the technical foundation for next-generation digital health interventions.
\section{Methods}

\subsection{COBRA Framework Architecture}

This section presents the COBRA framework, a comprehensive deep learning system designed for fine-grained classification of obesity-related behaviors using wrist-worn sensor data. This framework integrates advanced neural network architectures with behaviorally-oriented preprocessing pipelines to enable continuous, objective monitoring of energy balance-related behaviors essential for chronic obesity management.

\begin{figure}[h]
    \centering
    \includegraphics[width=0.9\textwidth]{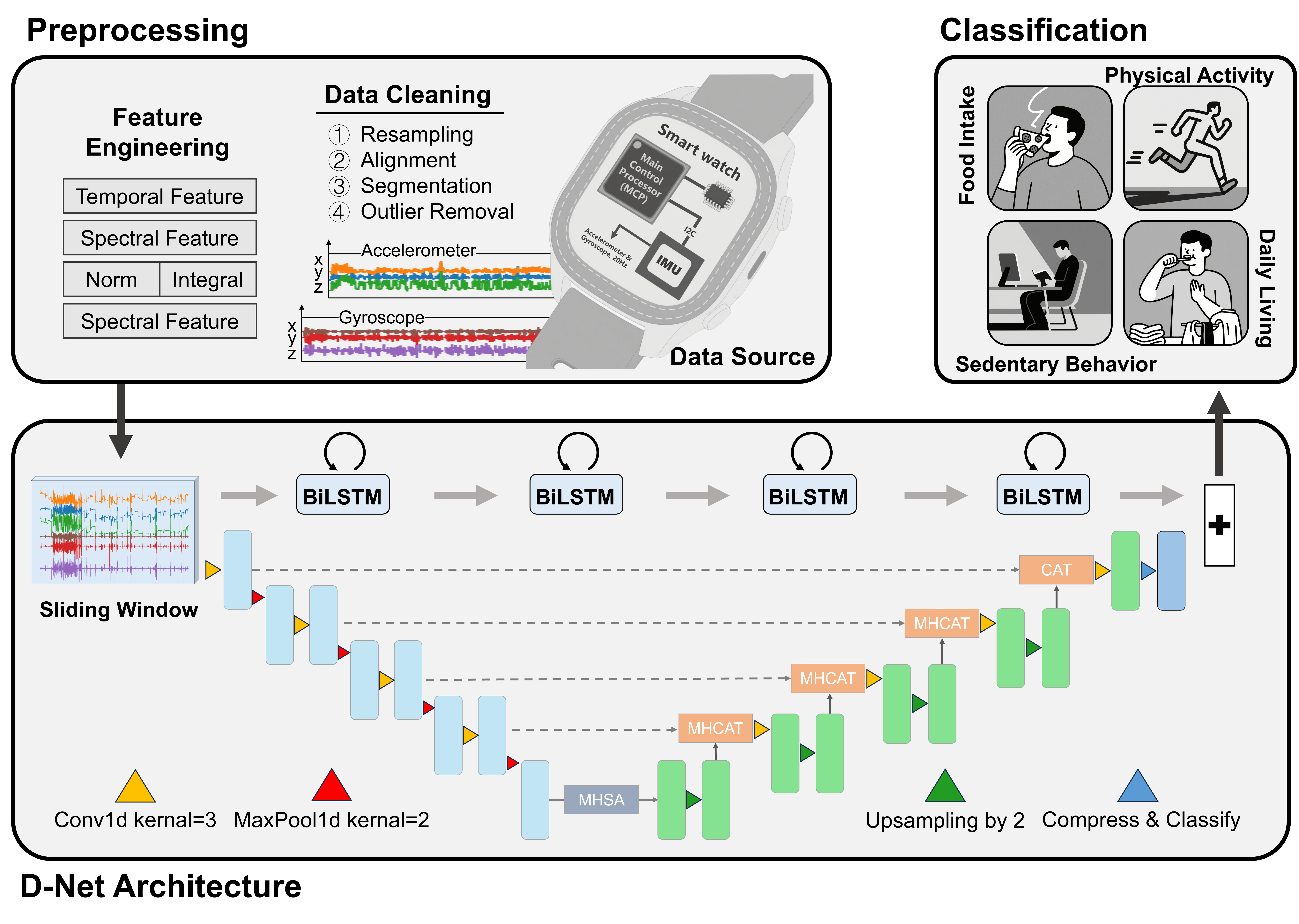}
    \caption{COBRA (Chronic Obesity Behavioral Recognition Architecture) framework overview. The system processes multimodal wrist-worn sensor data through specialized preprocessing, feature extraction, and hybrid neural network components to classify obesity-related behaviors into four behaviorally relevant categories.}
    \label{fig:COBRA}
\end{figure}

The COBRA framework employs a multi-stage processing pipeline that transforms raw wrist-worn sensor streams into behaviorally interpretable behavioral classifications. The architecture leverages the WISDM-Smart dataset \cite{misc_wisdmv2}, comprising synchronized accelerometer and gyroscope measurements collected during diverse daily activities. Data preprocessing begins with temporal alignment and interpolation to ensure synchronized measurements across sensor modalities, followed by sliding window segmentation with outlier detection and removal to maintain data integrity and reduce noise artifacts.

The feature extraction component of COBRA operates across both temporal and frequency domains to capture comprehensive behavioral signatures. Each temporal window undergoes Fast Fourier Transform (FFT) analysis and Power Spectral Density (PSD) computation to extract spectral characteristics, while acceleration magnitude norms and gyroscopic angular integrals provide motion-specific features. Raw sensor data standardization using z-score normalization enhances model generalizability and training stability \cite{gupta2021deep}.

The core classification engine employs a novel hybrid neural network architecture called D-Net, which synergizes a modified 23-layer U-Net backbone with multi-head self-attention mechanisms for spatial and cross-channel feature extraction. This architecture is complemented by a BiLSTM network that captures temporal dynamics and sequential dependencies. The fusion of these complementary pathways enables COBRA to simultaneously exploit localized activity patterns and long-range temporal information, resulting in superior performance across diverse behavioral categories.

COBRA's output classification scheme aligns with behaviorally relevant obesity management domains, categorizing daily activities into four primary categories: Food Intake, Physical Activity, Sedentary Behavior, and Daily Living. This hierarchical structure enables comprehensive behavioral profiling to support personalized obesity intervention strategies and long-term lifestyle modification programs.

\subsection{D-Net: Hybrid Deep Learning Architecture}

To address the complex requirements of chronic obesity behavioral monitoring, COBRA incorporates D-Net, a novel hybrid deep learning architecture that integrates multiple neural network paradigms for optimal feature extraction and classification performance. D-Net combines the spatial feature extraction capabilities of U-Net architectures with attention-based mechanisms and recurrent neural networks to process multivariate time-series data from wrist-worn sensors.

The foundation of D-Net builds upon the U-Net architecture \cite{chapter4pubevalzhang2019humanactivityrecognitionbasedonmotionsensorusingunet}, renowned for its ability to preserve spatial resolution while capturing hierarchical features through encoder-decoder pathways. Adapted specifically for multivariate time-series analysis, the U-Net component processes multi-channel inertial signals (3-axis accelerometer and 3-axis gyroscope) as structured temporal inputs, enabling dense activity classification at fine temporal resolutions.

\begin{figure}[!ht]
    \centering
    \begin{subfigure}[b]{0.48\textwidth}
        \includegraphics[width=\textwidth]{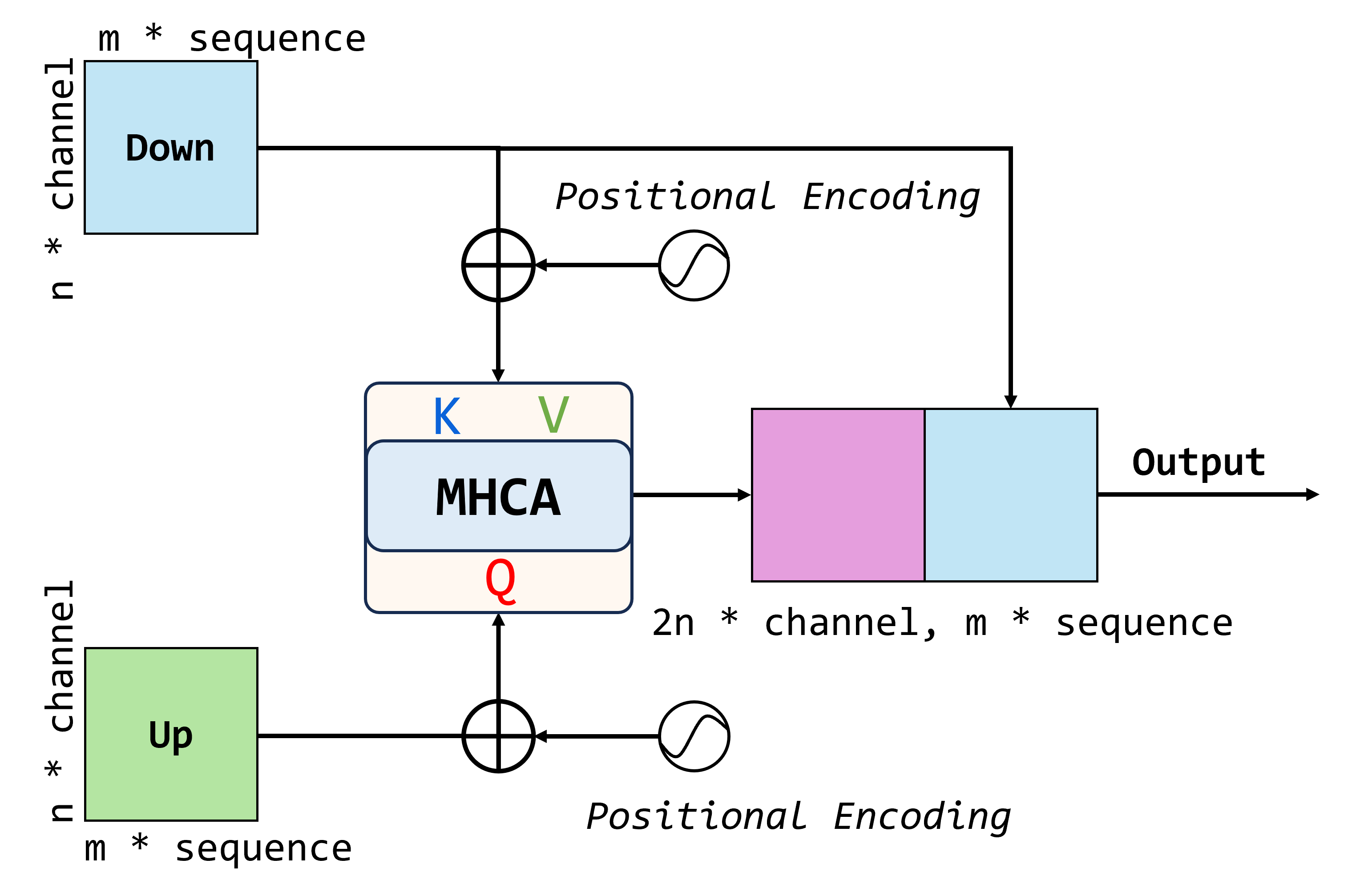}
        \caption{Multi-head cross-attention concatenation (MHCAT) module enabling feature alignment between encoder and decoder pathways in the D-Net architecture.}
        \label{fig:mhcat}
    \end{subfigure}
    \hfill
    \begin{subfigure}[b]{0.48\textwidth}
        \includegraphics[width=\textwidth]{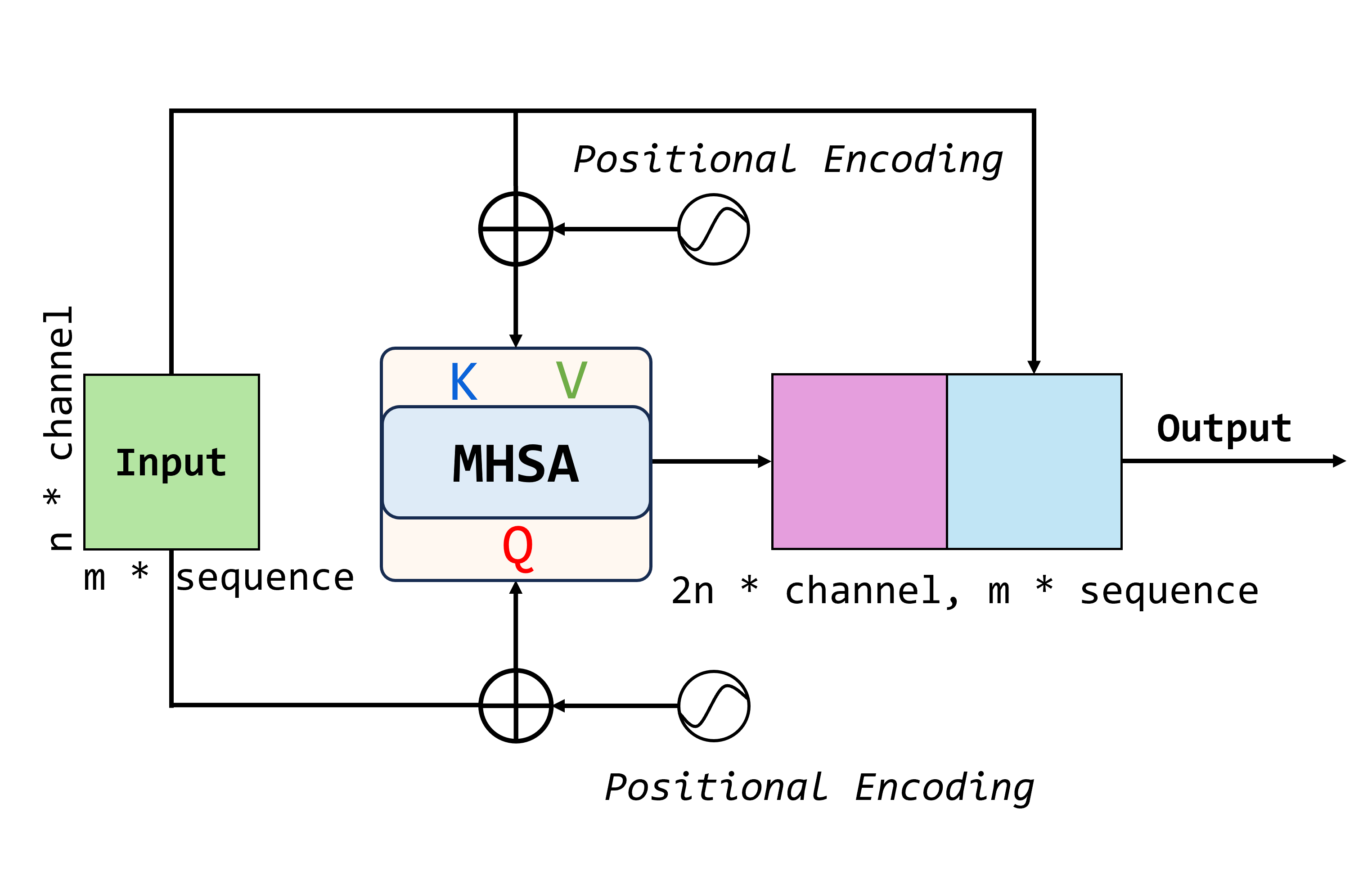}
        \caption{Multi-head self-attention (MHSA) module capturing temporal dependencies and cross-channel relationships in sensor data.}
        \label{fig:mhsa}
    \end{subfigure}
    \caption{Multi-head attention modules integrated within the COBRA framework's D-Net architecture for enhanced temporal and spatial feature extraction from wrist-worn sensor data.}
    \label{fig: MH_modules}
\end{figure}

D-Net enhances the base U-Net architecture through the integration of multi-head self-attention mechanisms positioned at strategic network locations. Multi-head self-attention modules are embedded at the bottleneck layer, while cross-attention mechanisms are distributed along the decoder pathway (Figure \ref{fig: MH_modules}). These attention components enable the model to dynamically focus on salient temporal features and establish alignments between low-level sensor patterns and high-level behavioral representations \cite{petit2021U-nettransformerSelfandcrossattentionformedicalimagesegmentation}. Such attention-based enhancements prove critical for distinguishing subtle yet behaviorally meaningful differences between similar behaviors, particularly for nuanced activities within the Food Intake category.

To capture long-term temporal dependencies inherent in human behavioral sequences, D-Net incorporates a parallel BiLSTM branch that processes the same input data through bidirectional recurrent analysis. The BiLSTM component excels at modeling sequential transitions and temporal patterns that extend beyond the local receptive fields of convolutional layers \cite{mekruksavanich2021deep}. Output features from both the attention-enhanced U-Net and BiLSTM pathways undergo fusion prior to final classification, enabling COBRA to integrate fine-grained spatial-temporal features with holistic sequential context.

By synergizing convolutional, attention-based, and recurrent neural network paradigms, D-Net addresses fundamental challenges in wearable-based activity recognition while providing a robust and flexible foundation for behaviorally meaningful behavioral assessment in COBRA's obesity management framework.

\subsection{Baseline Architectures for Performance Comparison}

To comprehensively evaluate COBRA's D-Net architecture, we implemented several state-of-the-art deep learning models representing different neural network paradigms for human activity recognition and time-series analysis. The baseline models include Multi-head Attention with 8 attention heads across 3 layers for capturing long-range dependencies \cite{vaswani2017attention}, Bidirectional LSTM with 128 hidden units per layer for sequential modeling \cite{graves2005framewisebilstm}, Convolutional BiLSTM combining 1D convolutions with recurrent processing \cite{mekruksavanich2021deep}, Transformer Encoder with positional encoding and 3 encoder blocks, Temporal Convolutional Network employing residual blocks with dilated causal convolutions \cite{bai2018empiricaltcn}, WaveNet utilizing dilated causal convolutions with gated activation mechanisms \cite{van2016wavenet}, and standard 1D U-Net with encoder-decoder pathways for hierarchical feature extraction \cite{ronneberger2015unet}.

These baseline architectures provide comprehensive coverage of major deep learning paradigms including attention-based mechanisms, recurrent neural networks, hybrid CNN-RNN approaches, and convolutional architectures, enabling thorough evaluation of COBRA's D-Net performance across different modeling approaches commonly employed in wearable sensor-based activity recognition systems.
\section{WISDM-Smart Dataset for COBRA Validation}

To evaluate the effectiveness and real-world applicability of the COBRA framework for chronic obesity behavioral monitoring, we conducted comprehensive validation using the publicly available \textbf{WISDM-Smart} dataset \cite{misc_wisdmv2}. This dataset provides a comprehensive source of multivariate time-series data from wrist-worn and pocket-based sensors, capturing diverse everyday behaviors essential for obesity management applications. The WISDM-Smart dataset was specifically selected for COBRA validation due to its unique comprehensive behavioral representation capability. Chronic obesity management is fundamentally a systematic intervention on daily life behavioral networks—dietary, exercise, and sedentary behaviors do not exist in isolation but form dynamically coupled systems (for example, office sitting duration affects post-meal energy consumption efficiency). The dataset captures this behavioral interconnectedness through multi-sensor fusion, and its coverage of multivariate time-series data (from active movement to fragmented daily activities) enables us to break through the limitations of traditional single behavior classification and construct realistic obesity management models.

For COBRA's validation objectives, we systematically reprocessed and reframed the WISDM-Smart dataset through an obesity management lens, reclassifying the original activity labels into behaviorally meaningful domains. This reclassification enables COBRA to provide directly actionable insights for obesity intervention strategies while maintaining compatibility with established activity recognition benchmarks.

\subsection{Dataset Characteristics and COBRA Integration}

The \textbf{WISDM Smartphone and Smartwatch Activity and Biometrics Dataset (WISDM-Smart)} represents a comprehensive publicly available resource designed for activity recognition and biometric analysis using mobile sensors \cite{misc_wisdmv2}. Developed by the WISDM Lab at Fordham University, the dataset comprises inertial sensor measurements from \textbf{51 subjects} performing \textbf{18 distinct activities} under controlled laboratory conditions. Each activity session was conducted for approximately \textbf{3 minutes}, resulting in over \textbf{15 million sensor measurements} sampled at \textbf{20 Hz} frequency.

The dataset includes synchronized recordings from both smartphone devices (positioned in participants' pockets) and LG G Watch smartwatches (worn on the dominant hand). For COBRA validation, we exclusively utilized smartwatch data to simulate realistic wrist-worn deployment scenarios that align with the framework's intended applications \cite{weiss2016actitracker}. This focus on wrist-worn sensors ensures that COBRA's performance evaluation reflects practical constraints and opportunities for long-term health monitoring in free-living conditions.

The WISDM-Smart dataset's activity repertoire encompasses a comprehensive range of behaviors spanning from dynamic physical activities to sedentary tasks, making it particularly suitable for evaluating COBRA's multi-category classification capabilities. The diversity of captured behaviors enables thorough assessment of the framework's ability to distinguish between subtle variations in movement patterns that are crucial for accurate obesity-related behavioral monitoring \cite{ross2009future}.

\subsection{Behavioral Reclassification for Obesity-Oriented COBRA Evaluation}

To align the WISDM-Smart dataset with COBRA's objectives for chronic obesity management, we systematically reclassified the original 18 activity labels into 4 behaviorally meaningful categories, as detailed in Table \ref{tab:behavior_classification}. This reclassification schema reflects the key behavioral domains that are central to evidence-based obesity interventions and long-term lifestyle modification programs \cite{bjsm2021health}. The reclassified dataset exhibits a balanced distribution with Physical Activity comprising 34.5\%, Food Intake 27.3\%, Sedentary Behavior 21.6\%, and Daily Living activities 16.4\% of the total behavioral instances.

\begin{table}[ht]
\centering
\caption{Behavioral classification schema implemented in COBRA for chronic obesity management. The four-category framework enables direct translation of recognition results into actionable insights for personalized intervention strategies.}
\setlength{\tabcolsep}{6pt}
\begin{tabular}{p{3.5cm}p{7cm}}
\toprule
\textbf{Behavioral Category} & \textbf{WISDM-Smart Activity Subcategories}\\
\midrule
Food Intake & Eating soup, Eating chips, Eating pasta, Drinking from cup, Eating sandwich\\
Physical Activity & Walking, Jogging, Stairs, Kicking, Playing catch, Dribbling\\
Sedentary Behavior & Sitting, Standing, Typing, Writing\\
Daily Living & Brushing teeth, Clapping, Folding clothes\\
\bottomrule
\end{tabular}
\label{tab:behavior_classification}
\end{table}

The \textbf{Food Intake} category encompasses various eating and drinking behaviors that directly contribute to energy intake monitoring, a critical component of obesity management \cite{thomaz2015practical,kyritsis2019modeling}. The \textbf{Physical Activity} category includes dynamic movements that contribute to energy expenditure, ranging from basic locomotion to structured exercise activities \cite{chen2003predicting}. \textbf{Sedentary Behavior} captures low-energy activities associated with metabolic risk and weight gain \cite{ross2009future, umbrella2024wearable}, while \textbf{Daily Living} represents routine activities that contribute to overall activity patterns but fall outside the primary energy balance domains.

\begin{figure}[!ht]
    \centering
    \includegraphics[width=0.8\textwidth]{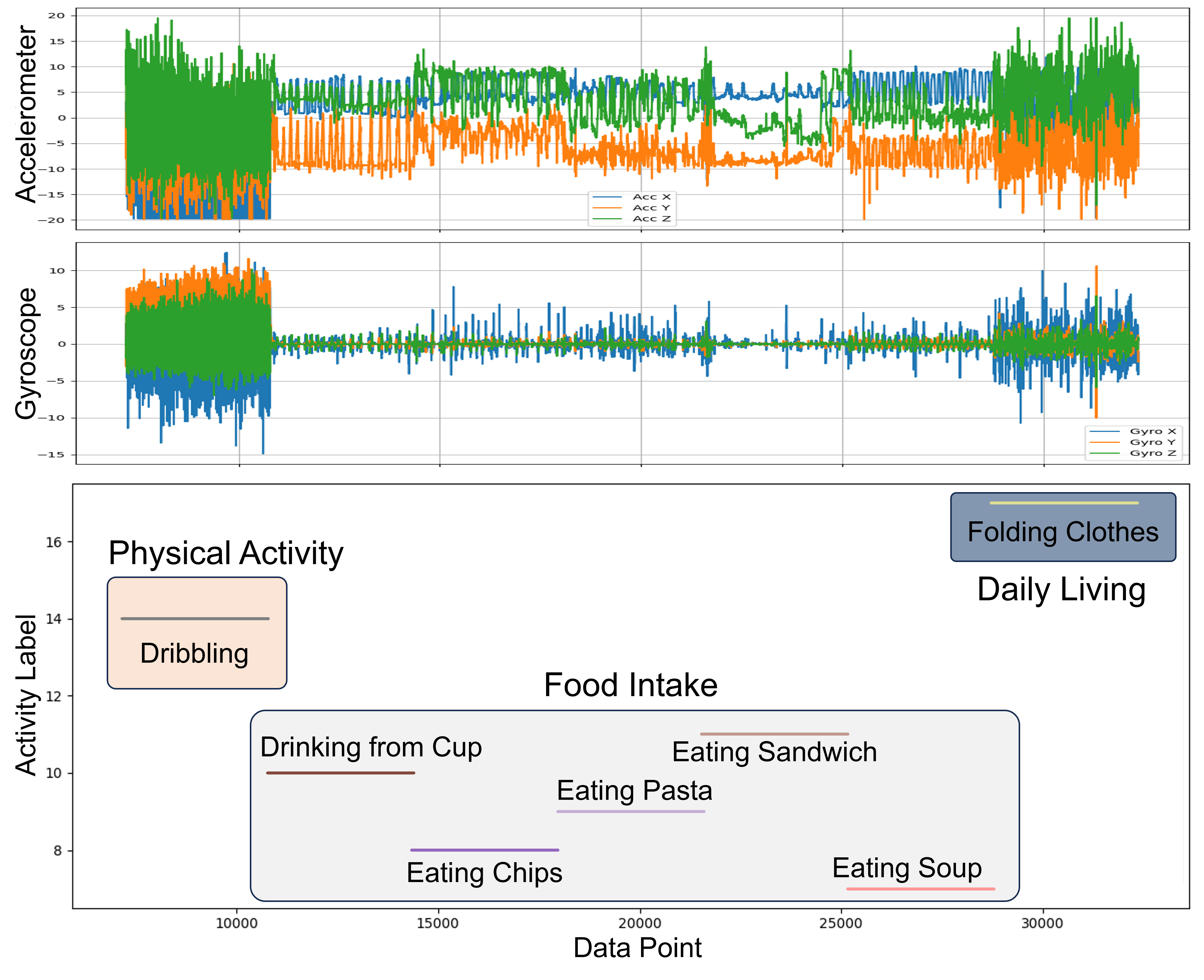}
    \caption{Multimodal sensor data visualization from WISDM-Smart dataset showing accelerometer readings (top), gyroscope measurements (middle), and corresponding behavioral category labels (bottom) for subject 1648. The data demonstrates the diverse signal patterns captured across COBRA's four behavioral categories, with distinct motion signatures for Food Intake, Physical Activity, Sedentary Behavior, and Daily Living activities.}
    \label{fig:WISDMv2_vis_session1}
\end{figure}

This behavioral reclassification enables COBRA to provide interpretable and directly actionable results that align with evidence-based obesity intervention protocols. The four-category framework supports real-time behavioral feedback, pattern analysis for personalized recommendations, and longitudinal monitoring of lifestyle changes during obesity treatment programs, as illustrated in Figure \ref{fig:WISDMv2_vis_session1}.

\subsection{Data Preprocessing Pipeline for COBRA Implementation}

Given the asynchronous nature of raw sensor streams and potential temporal misalignments in the WISDM-Smart dataset, COBRA implements a comprehensive preprocessing pipeline to ensure temporal consistency and label integrity across all behavioral categories. The pipeline addresses several data quality challenges inherent in real-world sensor deployments while maintaining compatibility with COBRA's behavioral classification objectives.

The preprocessing workflow begins with raw data ingestion and systematic label mapping, where accelerometer and gyroscope recordings are parsed and synchronized. Activity labels undergo standardization through COBRA's behavioral mapping function to align with the four-category behavioral schema, with numeric identifiers assigned for computational processing. Quality assurance procedures identify and address problematic data segments, including null or invalid labels, sensor measurement spikes, and temporal misalignments between accelerometer and gyroscope streams.

A critical preprocessing challenge addressed by COBRA involves timestamp irregularities in the nominally 20 Hz sampled data, which can significantly impact model performance if left unaddressed \cite{abdel2020st}. To mitigate these temporal inconsistencies, COBRA's preprocessing pipeline resamples all sensor signals to a standardized 50 Hz frequency, applies linear interpolation for missing timestamps, and trims activity segments to ensure synchronized durations between accelerometer and gyroscope measurements.

\subsection{Feature Extraction Strategy for COBRA's Multimodal Analysis}

Feature extraction represents a critical component of COBRA's preprocessing pipeline, designed to enhance the framework's ability to identify subtle behavioral patterns relevant to chronic obesity monitoring. The feature extraction strategy employs both frequency-domain analysis and time-domain signal processing to capture comprehensive motion characteristics from wrist-worn sensor data \cite{ellis2014random}.

COBRA's feature extraction pipeline implements two primary spectral analysis techniques: the \textbf{Fast Fourier Transform (FFT)} and \textbf{Power Spectral Density (PSD)} estimation \cite{proakis2007digital}. FFT decomposes time-domain signals into their constituent frequency components, computed as $X(f) = \sum_{t=0}^{N-1} x(t) \cdot e^{-i 2 \pi f t / N}$, revealing dominant frequencies characteristic of different movement patterns. PSD extends this analysis from an energy perspective, quantifying power distribution across frequencies as $\text{PSD}(f) = |X(f)|^2 / T$, where $T$ represents the observation window duration. These complementary methods enable the framework to identify frequency-domain characteristics that distinguish between different behavioral categories, particularly important for recognizing repetitive patterns in eating behaviors and physical activities.

To complement spectral analysis, COBRA incorporates two derived time-domain features that capture essential motion dynamics for obesity-related behavioral recognition. The \textbf{acceleration magnitude norm} is computed as $\|a(t)\| = \sqrt{a_x^2 + a_y^2 + a_z^2}$, providing orientation-independent measures of movement intensity valuable for distinguishing between sedentary and active behaviors. The \textbf{gyroscopic angular integration} feature estimates cumulative angular displacement through trapezoidal integration, capturing rotational movement patterns characteristic of specific behavioral categories.

The comprehensive feature extraction process generates a multi-dimensional feature matrix with dimensions (window size × 16 channels), combining FFT and PSD features (6 channels), acceleration magnitude norm (1 channel), gyroscopic angular integration (3 channels), and z-score normalized raw sensor data (6 channels). This enriched feature representation preserves the discriminative characteristics essential for accurate behavioral classification in obesity management applications while providing optimal compatibility with COBRA's D-Net architecture.
\section{COBRA Framework Evaluation and Results}

This section presents a comprehensive evaluation of the COBRA framework's performance for chronic obesity behavioral monitoring using the WISDM-Smart dataset. All experiments were conducted using exclusively smartwatch accelerometer and gyroscope data to simulate realistic wrist-worn deployment scenarios. Input data was segmented using a sliding window approach with a length of 200 samples (equivalent to 4 seconds at 50 Hz) and 75\% overlap between successive windows to ensure comprehensive temporal coverage.

The experimental design employed an 80-20 split for training and validation, with rigorous generalization assessment through leave-one-subject-out cross-validation on subjects 1648, 1649, and 1650. These subjects' data were completely excluded from the training phase to evaluate COBRA's ability to generalize to unseen individuals, a critical requirement for deployment in diverse patient populations.

Model training utilized the Adam optimizer with a learning rate of 0.001, batch size of 128, and categorical cross-entropy loss function, parameters selected to optimize COBRA's performance across the four obesity-related behavioral categories.

\subsection{COBRA Feature Engineering Strategy Evaluation}

To identify the optimal preprocessing pipeline for COBRA's objectives, we conducted a systematic evaluation of multiple feature extraction and normalization strategies. This analysis was essential for establishing COBRA's preprocessing foundation and ensuring robust performance across diverse obesity-related behaviors.

\begin{table}[!ht]
\centering
\caption{Feature extraction configurations evaluated for COBRA framework optimization. Each option represents a different combination of preprocessing steps designed to enhance behavioral pattern recognition for chronic obesity management.}
\label{tab:options}
\setlength{\tabcolsep}{6pt}
\begin{tabular}{c|@{\hspace{10pt}}c@{\hspace{10pt}}|c}
\toprule
Option & COBRA Preprocessing Strategy & Total Channels \\
\midrule
0 & Raw sensor data (no processing) & 6 \\
1 & Z-score normalization only & 6 \\
2 & Full features (FFT, PSD, norms, integrals) & 10 \\
3 & Raw data + full features & 16 \\
4 & Full features + z-score normalization & 16 \\
5 & Z-score normalization + full features & 16 \\
6 & Raw data + acceleration norm + gyro integration & 10 \\
7 & Z-score + acceleration norm + gyro integration & 10 \\
\bottomrule
\end{tabular}
\end{table}

The evaluation reveals that applying feature extraction followed by z-score normalization (Option 4) yields superior performance across architectures. This approach achieves the highest stability and accuracy by first extracting comprehensive spectral and temporal features (FFT, PSD, acceleration norms, and gyroscopic integrals) and subsequently applying z-score normalization, preserving discriminative information essential for obesity-related behavioral classification while ensuring numerical stability during training.

The analysis reveals that normalization-only approaches (Option 1) significantly degraded COBRA's performance compared to raw data processing, as z-score normalization discards absolute magnitude information essential for distinguishing between different obesity-related behaviors. Options 5 and 7, where normalization precedes feature extraction, perform particularly poorly as this approach disrupts the framework's ability to extract meaningful spectral-temporal characteristics. Interestingly, the combination of raw and extracted features (Option 3) did not substantially outperform raw data alone, suggesting that COBRA's D-Net architecture possesses sufficient intrinsic capacity to learn meaningful representations directly from raw sensor data.

To ensure that Option 4's superior performance was not architecture-specific, we validated its effectiveness across multiple deep learning models representative of different paradigms commonly used in activity recognition. Table~\ref{tab:option4_2} presents average F1 scores achieved by three diverse architectures—Attention, Temporal Convolutional Network (TCN), and U-Net—across all preprocessing options.

\begin{table}[!ht]
\centering
\caption{COBRA framework performance comparison across different neural network architectures and feature extraction strategies. Option 4 demonstrates consistent superiority across diverse model types, validating its selection as COBRA's optimal preprocessing approach for chronic obesity behavioral monitoring.}
\label{tab:option4_2}
\setlength{\tabcolsep}{6pt}
\begin{tabular}{c|c|c|c|c}
\hline
\textbf{Option} & \textbf{Attention} & \textbf{TCN} & \textbf{U-Net} & \textbf{Average} \\ \hline
0 & 0.742 & 0.801 & 0.761 & 0.768 \\
1 & 0.654 & 0.775 & 0.749 & 0.726 \\
2 & 0.770 & 0.805 & 0.779 & 0.785 \\
3 & \textbf{0.784} & 0.776 & 0.771 & 0.777 \\
\textbf{4} & 0.773 & \textbf{0.811} & \textbf{0.809} & \textbf{0.797} \\
5 & 0.762 & 0.794 & 0.746 & 0.767 \\
6 & 0.768 & 0.786 & 0.797 & 0.784 \\
7 & 0.719 & 0.806 & 0.775 & 0.767 \\
\hline
\end{tabular}
\end{table}

Option 4 achieved the highest average F1 score (0.797) across all three neural network architectures, confirming its robustness and architecture-agnostic superiority for COBRA applications. While Option 3 yielded marginally better results on the Attention model, Option 4 demonstrated more balanced and consistent performance, ranking either first or second across all architectures tested. Based on these comprehensive evaluations, Option 4 was adopted as COBRA's standard preprocessing strategy for all subsequent experiments.

\subsection{COBRA Performance Comparison with State-of-the-Art Models}

This section presents a comprehensive performance comparison between COBRA's D-Net architecture and established deep learning baselines for obesity-related behavioral classification. The evaluation focuses specifically on COBRA's four-category classification scheme designed for chronic obesity management applications.

\begin{table}[h]
\centering
\caption{COBRA framework performance comparison with baseline deep learning models for four-category obesity-related behavioral classification. D-Net demonstrates superior performance across all evaluation metrics, establishing COBRA as a robust solution for chronic obesity behavioral monitoring.}
\label{tab:model_performance4}
\setlength{\tabcolsep}{6pt}
\begin{tabular}{l@{\hspace{10pt}}cccc}
\toprule
\textbf{Model} & \textbf{Accuracy} & \textbf{Macro Precision} & \textbf{Macro Recall} & \textbf{Macro F1} \\
\midrule
Attention    & 0.9427 & 0.9415 & 0.9387 & 0.9395 \\
BiLSTM       & 0.9568 & 0.9545 & 0.9567 & 0.9555 \\
CNNBiLSTM    & 0.9470 & 0.9463 & 0.9456 & 0.9449 \\
Encoder      & 0.3384 & 0.0846 & 0.2500 & 0.1264 \\
TCN          & 0.9514 & 0.9494 & 0.9507 & 0.9499 \\
WaveNet      & 0.9481 & 0.9456 & 0.9444 & 0.9446 \\
U-Net         & 0.9535 & 0.9507 & 0.9556 & 0.9527 \\
\textbf{D-Net (Ours)}        & \textbf{0.9686} & \textbf{0.9687} & \textbf{0.9684} & \textbf{0.9685} \\
\bottomrule
\end{tabular}
\end{table}

Table~\ref{tab:model_performance4} demonstrates COBRA's superior performance across all evaluation metrics for the four obesity-related behavioral categories: Physical Activity, Food Intake, Sedentary Behavior, and Daily Living. COBRA's D-Net architecture achieves the highest scores in all metrics: 96.86\% accuracy, 96.87\% macro precision, 96.84\% macro recall, and 96.85\% macro F1-score.

These results represent significant improvements over strong baseline architectures commonly used in activity recognition. COBRA outperforms the second-best model (BiLSTM) by 1.18\% in accuracy and 1.30\% in macro F1-score. More importantly, COBRA demonstrates consistent excellence across all metrics, indicating balanced performance across the four behavioral categories rather than excelling in some categories while underperforming in others.

The substantial performance gap between COBRA and simpler architectures like the basic Attention model (2.56\% accuracy improvement) and CNNBiLSTM (2.16\% improvement) highlights the effectiveness of COBRA's hybrid design combining U-Net spatial modeling, attention mechanisms, and BiLSTM temporal processing. This architecture proves particularly well-suited for capturing the complex patterns inherent in obesity-related behavioral monitoring.

For chronic obesity management applications, COBRA's outstanding performance translates directly into more reliable behavioral monitoring, reduced classification errors that could lead to inappropriate intervention recommendations, and potentially increased effectiveness and confidence in automated dietary and activity tracking systems deployed in health monitoring settings.

\begin{figure}[!ht]
    \centering
    \includegraphics[width=0.6\textwidth]{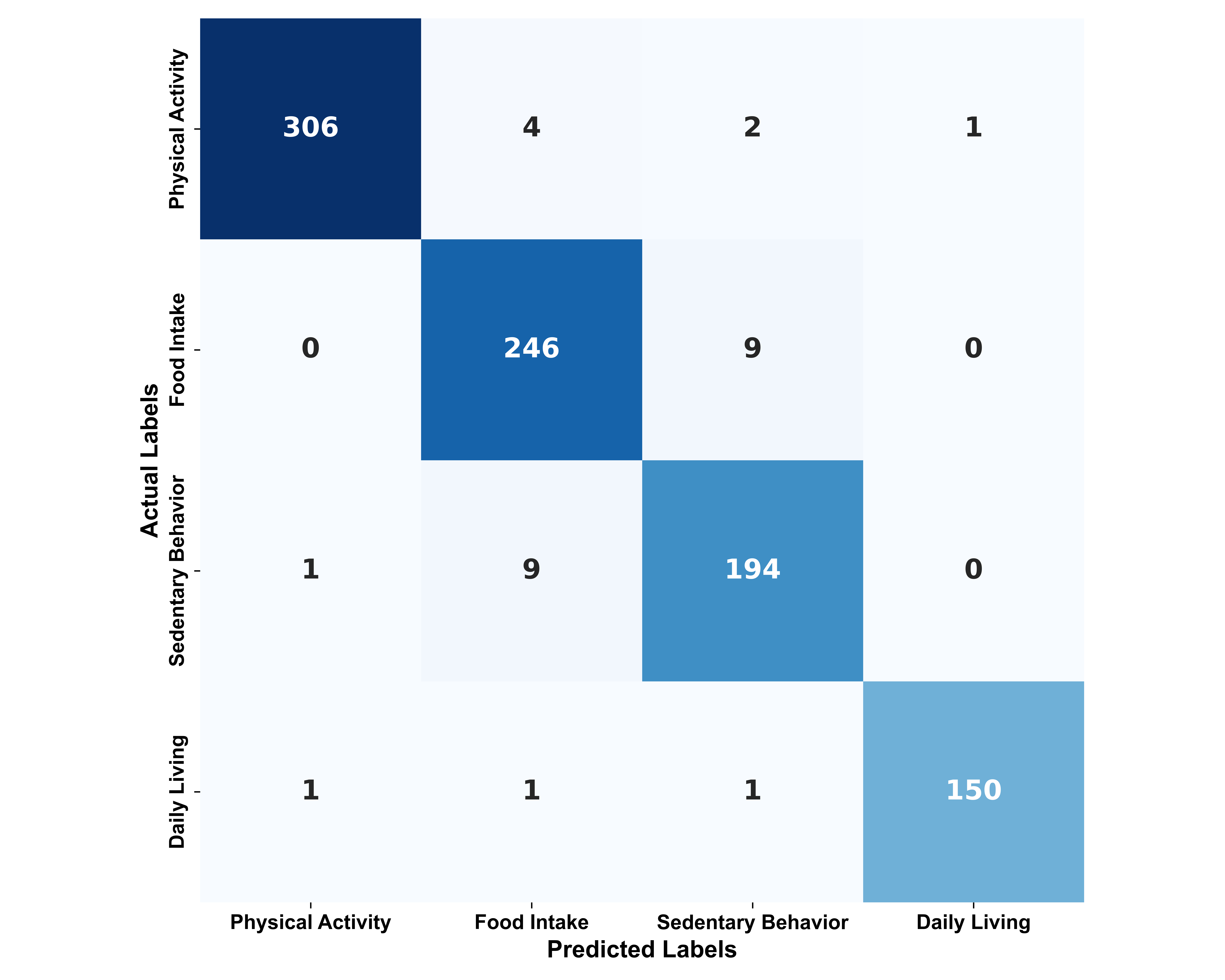}
    \caption{COBRA D-Net confusion matrix for four-category obesity-related behavioral classification. The matrix demonstrates excellent classification performance across all categories relevant to chronic obesity management, with minimal misclassification between Food Intake, Physical Activity, Sedentary Behavior, and Daily Living activities. Strong diagonal dominance indicates COBRA's reliability for future deployment.}
    \label{fig:heatmap4}
\end{figure}

Figure~\ref{fig:heatmap4} provides detailed insights into COBRA's classification performance through a confusion matrix analysis. The results demonstrate excellent discrimination across all four obesity-related behavioral categories, with strong diagonal dominance indicating accurate classification and minimal inter-category confusion.

Specifically, COBRA correctly classifies 306 instances of Physical Activity, 246 instances of Food Intake, 194 instances of Sedentary Behavior, and 150 instances of Daily Living activities, achieving category-specific F1-scores of 98.55\%, 95.53\%, 94.63\%, and 98.68\% respectively. The minimal off-diagonal elements indicate that COBRA rarely confuses behaviors from different categories, a critical requirement for reliable obesity monitoring where misclassification between food intake and other activities could significantly impact dietary tracking accuracy.

From an obesity management perspective, COBRA's performance enables several key applications. The high accuracy in Food Intake classification with 246 correct predictions and minimal false positives enables reliable automated dietary monitoring, addressing a major limitation of self-reported food diaries that are prone to underreporting and recall bias. Accurate Physical Activity classification supports objective measurement of energy expenditure and exercise adherence, enabling healthcare providers to assess patient compliance with prescribed activity levels and adjust intervention strategies accordingly. Precise identification of Sedentary Behavior provides insights into prolonged inactivity patterns associated with metabolic dysfunction and weight gain, enabling targeted interventions to reduce sedentary time. The combined classification across all four categories enables holistic behavioral profiling that supports personalized obesity intervention strategies based on individual activity patterns and lifestyle characteristics.
\section{Discussion and Conclusion}

This study introduces COBRA (Chronic Obesity Behavioral Recognition Architecture), a comprehensive deep learning framework for classifying obesity-related behaviors using wrist-worn sensor data. Beyond the technical achievements of the D-Net model, which achieved 96.86\% accuracy with F1-scores exceeding 94\% across all four behaviorally relevant categories, COBRA's architectural design demonstrates the potential for practical deployment in nuanced obesity management interventions. The framework's four-category behavioral classification (Food Intake, Physical Activity, Sedentary Behavior, Daily Living) provides the foundation for future clinical continuous energy balance monitoring researches, where daily energy intake and expenditure could be automatically tracked and analyzed.

COBRA's demonstrated accuracy in detecting energy intake behaviors addresses a longstanding limitation in obesity care: the unreliability of self-reported dietary logs. By offering an objective alternative, the framework has the potential to improve dietary monitoring without burdening users. The comprehensive behavioral classification enables healthcare providers to assess multiple dimensions of energy balance simultaneously, supporting evidence-based obesity interventions that address both energy intake and expenditure components \cite{bjsm2021health}. Building upon these technical capabilities, COBRA's potential application in obesity management suggests several promising intervention strategies based on fine-grainted behavioral pattern recognition. Future implementations could trigger personalized interventions when energy expenditure falls below prescribed thresholds or when prolonged sedentary behavior is detected, such as activity reminders, step goal adjustments, or structured exercise recommendations \cite{cadmus2016technology,bjsm2021health}. Similarly, when eating frequency or duration patterns deviate from established baselines, the framework could facilitate just-in-time dietary interventions, portion control reminders, or mindful eating prompts \cite{thomaz2015practical,rahman2016predicting}.

Through extensive evaluation, we found that applying feature extraction followed by z-score normalization (Option 4) yields superior performance across architectures, achieving an average F1-score of 0.797. The integration of spectral-temporal features including FFT, PSD, acceleration norm, and gyroscopic integrals captures both periodicity and motion signatures, enabling robust recognition across diverse activity types. COBRA's D-Net hybrid architecture, combining U-Net's spatial modeling, attention mechanisms, and BiLSTM-based temporal learning, effectively captures both localized patterns and long-range dependencies essential for complex human activity recognition. The attention mechanisms prove particularly valuable for distinguishing between similar behavioral categories, dynamically focusing on salient temporal features while facilitating alignment between low-level sensor patterns and high-level behavioral representations.

COBRA demonstrates high generalizability, with minimal demographic performance variance (less than 3\%), supporting its fairness and potential for wide-scale adoption. Its wrist-based implementation enables continuous, passive monitoring, aligning with precision medicine goals and current wearable technology trends. However, deployment in real-world settings requires overcoming several challenges including sensor variability across device manufacturers, complex free-living activity patterns, and the need for integration with complementary technologies for nutritional granularity beyond behavioral classification. The framework's continuous monitoring capability suggests opportunities for healthcare providers to implement adaptive behavioral modification protocols, potentially adjusting intervention intensity based on individual compliance patterns and progress toward weight management goals \cite{umbrella2024wearable}.

COBRA represents a significant advancement in wearable-based behavioral monitoring, delivering high classification accuracy with robust preprocessing frameworks. Future work will involve real-world validation in clinical studies, sensor fusion with contextual data, and the development of adaptive intervention strategies leveraging rich behavioral signals for improved outcomes. Future integration with mobile health applications and clinical decision support systems could enable personalized feedback loops, where behavioral insights inform treatment plan modifications and support precision medicine approaches to obesity management \cite{weiss2016actitracker}. The framework's technology could extend beyond obesity management to eating disorder treatment, diabetes management, and elderly care programs. By enabling more precise behavioral monitoring with reduced reliance on self-reporting, COBRA could improve intervention effectiveness while reducing healthcare costs, making intensive monitoring programs more accessible to broader patient populations.

\printbibliography

@article{vaswani2017attention,
  title={Attention is all you need},
  author={Vaswani, Ashish and Shazeer, Noam and Parmar, Niki and Uszkoreit, Jakob and Jones, Llion and Gomez, Aidan N and Kaiser, {\L}ukasz and Polosukhin, Illia},
  journal={Advances in neural information processing systems},
  volume={30},
  year={2017}
}

@article{graves2005framewisebilstm,
  title={Framewise phoneme classification with bidirectional LSTM and other neural network architectures},
  author={Graves, Alex and Schmidhuber, J{\"u}rgen},
  journal={Neural networks},
  volume={18},
  number={5-6},
  pages={602--610},
  year={2005},
  publisher={Elsevier}
}

@article{bai2018empiricaltcn,
  title={An empirical evaluation of generic convolutional and recurrent networks for sequence modeling},
  author={Bai, Shaojie and Kolter, J Zico and Koltun, Vladlen},
  journal={arXiv preprint arXiv:1803.01271},
  year={2018}
}

@article{van2016wavenet,
  title={Wavenet: A generative model for raw audio},
  author={Van Den Oord, Aaron and Dieleman, Sander and Zen, Heiga and Simonyan, Karen and Vinyals, Oriol and Graves, Alex and Kalchbrenner, Nal and Senior, Andrew and Kavukcuoglu, Koray and others},
  journal={arXiv preprint arXiv:1609.03499},
  volume={12},
  year={2016}
}

@inproceedings{ronneberger2015unet,
  title={U-net: Convolutional networks for biomedical image segmentation},
  author={Ronneberger, Olaf and Fischer, Philipp and Brox, Thomas},
  booktitle={Medical image computing and computer-assisted intervention--MICCAI 2015: 18th international conference, Munich, Germany, October 5-9, 2015, proceedings, part III 18},
  pages={234--241},
  year={2015},
  organization={Springer}
}

@article{ross2009future,
  title={The future of obesity reduction: beyond weight loss},
  author={Ross, Robert and Bradshaw, Alison J},
  journal={Nature Reviews Endocrinology},
  volume={5},
  number={6},
  pages={319--325},
  year={2009},
  publisher={Nature Publishing Group UK London}
}

@book{proakis2007digital,
  title={Digital signal processing: principles, algorithms, and applications, 4/E},
  author={Proakis, John G},
  year={2007},
  publisher={Pearson Education India}
}

@article{ElmalehSachs2023,
  title={Obesity management in adults: a review},
  author={Elmaleh-Sachs, Ariela and Schwartz, Jessica L. and Bramante, Carolyn T. and Nicklas, Jacinda M. and Gudzune, Kimberly A. and Jay, Melanie},
  journal={JAMA},
  volume={330},
  number={20},
  pages={2000--2015},
  year={2023},
  doi={10.1001/jama.2023.2000}
}

@online{WHOobesity2022website,
  title={Obesity and overweight},
  author={{World Health Organization}},
  year={2022},
  url={https://www.who.int/news-room/fact-sheets/detail/obesity-and-overweight},
  note={Accessed: 2025-05-11}
}

@article{chapter4pubevalzhang2019humanactivityrecognitionbasedonmotionsensorusingunet,
  title={Human activity recognition based on motion sensor using u-net},
  author={Zhang, Yong and Zhang, Zhao and Zhang, Yu and Bao, Jie and Zhang, Yifan and Deng, Haiqin},
  journal={IEEE access},
  volume={7},
  pages={75213--75226},
  year={2019},
  publisher={IEEE}
}

@inproceedings{petit2021U-nettransformerSelfandcrossattentionformedicalimagesegmentation,
  title={U-net transformer: Self and cross attention for medical image segmentation},
  author={Petit, Olivier and Thome, Nicolas and Rambour, Clement and Themyr, Loic and Collins, Toby and Soler, Luc},
  booktitle={Machine Learning in Medical Imaging: 12th International Workshop, MLMI 2021, Held in Conjunction with MICCAI 2021, Strasbourg, France, September 27, 2021, Proceedings 12},
  pages={267--276},
  year={2021},
  organization={Springer}
}

@misc{misc_wisdmv2,
  author       = {Weiss,Gary},
  title        = {{WISDM Smartphone and Smartwatch Activity and Biometrics Dataset }},
  year         = {2019},
  howpublished = {UCI Machine Learning Repository},
  note         = {{DOI}: https://doi.org/10.24432/C5HK59}
}

@inproceedings{ramachandran2020evaluation,
 title={Evaluation of feature engineering on wearable sensor-based fall detection},
 author={Ramachandran, Anita and Ramesh, Adarsh and Karuppiah, Anupama},
 booktitle={2020 International Conference on Information Networking (ICOIN)},
 pages={110--114},
 year={2020},
 organization={IEEE}
}

@inproceedings{mondol2020hawad,
 title={HAWAD: Hand washing detection using wrist wearable inertial sensors},
 author={Mondol, Md Abu Sayeed and Stankovic, John A},
 booktitle={2020 16th International Conference on Distributed Computing in Sensor Systems (DCOSS)},
 pages={11--18},
 year={2020},
 organization={IEEE}
}

@inproceedings{mekruksavanich2022deep,
 title={Deep learning models for daily living activity recognition based on wearable inertial sensors},
 author={Mekruksavanich, Sakorn and Jantawong, Ponnipa and Hnoohom, Narit and Jitpattanakul, Anuchit},
 booktitle={2022 19th International Joint Conference on Computer Science and Software Engineering (JCSSE)},
 pages={1--5},
 year={2022},
 organization={IEEE}
}

@article{gupta2021deep,
 title={Deep learning based human activity recognition (HAR) using wearable sensor data},
 author={Gupta, Saurabh},
 journal={International Journal of Information Management Data Insights},
 volume={1},
 number={2},
 pages={100046},
 year={2021},
 publisher={Elsevier}
}

@article{mekruksavanich2021deep,
 title={Deep convolutional neural network with rnns for complex activity recognition using wrist-worn wearable sensor data},
 author={Mekruksavanich, Sakorn and Jitpattanakul, Anuchit},
 journal={Electronics},
 volume={10},
 number={14},
 pages={1685},
 year={2021},
 publisher={MDPI}
}

@article{khatun2022deep,
 title={Deep CNN-LSTM with self-attention model for human activity recognition using wearable sensor},
 author={Khatun, Mst Alema and Yousuf, Mohammad Abu and Ahmed, Sabbir and Uddin, Md Zia and Alyami, Salem A and Al-Ashhab, Samer and Akhdar, Hanan F and Khan, Asaduzzaman and Azad, Akm and Moni, Mohammad Ali},
 journal={IEEE Journal of Translational Engineering in Health and Medicine},
 volume={10},
 pages={1--16},
 year={2022},
 publisher={IEEE}
}

@article{dirgova2022wearable,
 title={Wearable sensor-based human activity recognition with transformer model},
 author={Dirgov{'a} Lupt{'a}kov{'a}, Iveta and Kubov{\v{c}}{'\i}k, Martin and Posp{'\i}chal, Ji{\v{r}}{'\i}},
 journal={Sensors},
 volume={22},
 number={5},
 pages={1911},
 year={2022},
 publisher={MDPI}
}

@article{li2022human,
 title={Human activity recognition based on residual network and BiLSTM},
 author={Li, Yong and Wang, Luping},
 journal={Sensors},
 volume={22},
 number={2},
 pages={635},
 year={2022},
 publisher={MDPI}
}

@article{mim2023gru,
 title={GRU-INC: An inception-attention based approach using GRU for human activity recognition},
 author={Mim, Taima Rahman and Amatullah, Maliha and Afreen, Sadia and Yousuf, Mohammad Abu and Uddin, Shahadat and Alyami, Salem A and Hasan, Khondokar Fida and Moni, Mohammad Ali},
 journal={Expert Systems with Applications},
 volume={216},
 pages={119419},
 year={2023},
 publisher={Elsevier}
}

@inproceedings{weiss2016actitracker,
 title={Actitracker: a smartphone-based activity recognition system for improving health and well-being},
 author={Weiss, Gary M and Lockhart, Jeffrey W and Pulickal, Tony T and McHugh, Paul T and Ronan, Isaac H and Timko, Jessica L},
 booktitle={2016 IEEE international conference on data science and advanced analytics (DSAA)},
 pages={682--688},
 year={2016},
 organization={IEEE}
}

@inproceedings{zhu2015using,
 title={Using deep learning for energy expenditure estimation with wearable sensors},
 author={Zhu, Jindan and Pande, Amit and Mohapatra, Prasant and Han, Jay J},
 booktitle={2015 17th International Conference on E-health Networking, Application \& Services (HealthCom)},
 pages={501--506},
 year={2015},
 organization={IEEE}
}

@article{ellis2014random,
 title={A random forest classifier for the prediction of energy expenditure and type of physical activity from wrist and hip accelerometers},
 author={Ellis, Katherine and Kerr, Jacqueline and Godbole, Suneeta and Lanckriet, Gert and Wing, David and Marshall, Simon},
 journal={Physiological measurement},
 volume={35},
 number={11},
 pages={2191},
 year={2014},
 publisher={IOP Publishing}
}

@article{chen2003predicting,
 title={Predicting energy expenditure of physical activity using hip-and wrist-worn accelerometers},
 author={Chen, Kong Y and Acra, Sari A and Majchrzak, Karen and Donahue, Candice L and Baker, Lemont and Clemens, Linda and Sun, Ming and Buchowski, Maciej S},
 journal={Diabetes technology \& therapeutics},
 volume={5},
 number={6},
 pages={1023--1033},
 year={2003},
 publisher={Mary Ann Liebert, Inc.}
}

@article{biscuit2017shortterm,
 title={Short-term efficacy of an innovative mobile phone technology-based intervention for weight management for overweight and obese adolescents: pilot study},
 author={Chen, Jyu-Lin and Guedes, Claudia M and Cooper, Bruce A and Lung, Audrey E and others},
 journal={Interactive journal of medical research},
 volume={6},
 number={2},
 pages={e7860},
 year={2017},
 publisher={JMIR Publications Inc., Toronto, Canada}
}

@article{cadmus2016technology,
 title={Technology-and phone-based weight loss intervention: pilot RCT in women at elevated breast cancer risk},
 author={Hartman, Sheri J and Nelson, Sandahl H and Cadmus-Bertram, Lisa A and Patterson, Ruth E and Parker, Barbara A and Pierce, John P},
 journal={American journal of preventive medicine},
 volume={51},
 number={5},
 pages={714--721},
 year={2016},
 publisher={Elsevier}
}

@article{umbrella2024wearable,
 title={Wearable devices to improve physical activity and reduce sedentary behaviour: an umbrella review},
 author={Longhini, Jessica and Marzaro, Chiara and Bargeri, Silvia and Palese, Alvisa and Dell’Isola, Andrea and Turolla, Andrea and Pillastrini, Paolo and Battista, Simone and Castellini, Greta and Cook, Chad and others},
 journal={Sports Medicine-Open},
 volume={10},
 number={1},
 pages={9},
 year={2024},
 publisher={Springer}
}

@article{bjsm2021health,
 title={Health wearable devices for weight and BMI reduction in individuals with overweight/obesity and chronic comorbidities: systematic review and network meta-analysis},
 author={McDonough, Daniel J and Su, Xiwen and Gao, Zan},
 journal={British journal of sports medicine},
 volume={55},
 number={16},
 pages={917--925},
 year={2021},
 publisher={BMJ Publishing Group Ltd and British Association of Sport and Exercise Medicine}
}

@article{abdel2020st,
 title={ST-DeepHAR: Deep learning model for human activity recognition in IoHT applications},
 author={Abdel-Basset, Mohamed and Hawash, Hossam and Chakrabortty, Ripon K and Ryan, Michael and Elhoseny, Mohamed and Song, Houbing},
 journal={IEEE Internet of Things Journal},
 volume={8},
 number={6},
 pages={4969--4979},
 year={2020},
 publisher={IEEE}
}

@article{dong2013detecting,
 title={Detecting periods of eating during free-living by tracking wrist motion},
 author={Dong, Yujie and Scisco, Jenna and Wilson, Mike and Muth, Eric and Hoover, Adam},
 journal={IEEE journal of biomedical and health informatics},
 volume={18},
 number={4},
 pages={1253--1260},
 year={2013},
 publisher={IEEE}
}

@inproceedings{thomaz2015practical,
 title={A practical approach for recognizing eating moments with wrist-mounted inertial sensing},
 author={Thomaz, Edison and Essa, Irfan and Abowd, Gregory D},
 booktitle={Proceedings of the 2015 ACM international joint conference on pervasive and ubiquitous computing},
 pages={1029--1040},
 year={2015}
}

@ARTICLE{kyritsis2019modeling,
 author={Kyritsis, Konstantinos and Diou, Christos and Delopoulos, Anastasios},
 journal={IEEE Journal of Biomedical and Health Informatics}, 
 title={Modeling Wrist Micromovements to Measure In-Meal Eating Behavior From Inertial Sensor Data}, 
 year={2019},
 volume={23},
 number={6},
 pages={2325-2334},
 doi={10.1109/JBHI.2019.2892011}
}

@article{mirtchouk2017recognizing,
 title={Recognizing eating from body-worn sensors: Combining free-living and laboratory data},
 author={Mirtchouk, Mark and Lustig, Drew and Smith, Alexandra and Ching, Ivan and Zheng, Min and Kleinberg, Samantha},
 journal={Proceedings of the ACM on Interactive, Mobile, Wearable and Ubiquitous Technologies},
 volume={1},
 number={3},
 pages={1--20},
 year={2017},
 publisher={ACM New York, NY, USA}
}

@inproceedings{rahman2016predicting,
 title={Predicting" about-to-eat" moments for just-in-time eating intervention},
 author={Rahman, Tauhidur and Czerwinski, Mary and Gilad-Bachrach, Ran and Johns, Paul},
 booktitle={Proceedings of the 6th International Conference on Digital Health Conference},
 pages={141--150},
 year={2016}
}

@article{farooq2016novel,
 title={A novel wearable device for food intake and physical activity recognition},
 author={Farooq, Muhammad and Sazonov, Edward},
 journal={Sensors},
 volume={16},
 number={7},
 pages={1067},
 year={2016},
 publisher={MDPI}
}

\end{document}